\documentclass[12pt]{article}
\usepackage[T1]{fontenc}
\usepackage[utf8]{inputenc}
\usepackage[brazil,english]{babel}
\usepackage{amssymb,amsmath,amsbsy}
\usepackage[colorlinks=true]{hyperref}
\usepackage{varioref}
\usepackage{graphicx}
    \graphicspath{{./Figures/}}
\usepackage{subfigure}
\usepackage{floatrow}
\usepackage{mathtools,stackengine}
\usepackage[tmargin=2cm,lmargin=2cm,rmargin=2cm,bmargin=2cm]{geometry}
\usepackage[square,numbers,sort&compress]{natbib}
\usepackage{longtable}
\usepackage{xcolor}
\usepackage{microtype}
\usepackage{multirow}
\usepackage{stix}
\usepackage{array}
\usepackage{caption}

\newcommand{\sqabs}[1]{{\left|#1\right|}^2}
\newcommand{\ZZ}{\mathbb{Z}}
\newcommand{\RR}{\mathbb{R}}
\newcommand{\CC}{\mathbb{C}}

\newcommand{\hypergeom}[1]{{}_2F_1\left(#1\right)}
\newcommand{\diff}[2]{\frac{\text{d}#1}{\text{d}#2}}
\newcommand{\ddiff}[2]{\frac{\text{d}^2#1}{{\text{d}#2}^2}}
\newcommand{\pdiff}[2]{\frac{\partial#1}{\partial#2}}
\newcommand{\pddiff}[2]{\frac{\partial^2#1}{\partial{#2}^2}}
\newcommand{\beq}{\begin{equation}}
\newcommand{\eeq}{\end{equation}}

\title{\bf{Energy eigenstates
 of position-dependent mass particles in a spherical quantum dot}}
\author{R. M. Lima* and H. R. Christiansen** \\
\small{*CBPF -- Centro Brasileiro de Pesquisas Físicas, CEP 22290-180, Rio de Janeiro, Brazil }  \\
\small{**IFCE -- Instituto Federal de Educação, Ciência e Tecnologia do Ceará, Maranguape, CEP 61940-750, Ceará, Brazil}
\\
\small{corresponding author: hugo.christiansen@ifce.edu.br}}
\date{}
\begin{document}
\maketitle
\begin{abstract}
We obtain the exact energy spectrum of nonuniform mass particles for a collection of Hamiltonians in a three-dimensional  approach to a quantum dot. By considering a set of generalized Schrödinger equations with different orderings between the particle's momentum and mass, the energy bound-states are calculated analytically for hard boundary conditions. The present results are of interest in atomic physics and quantum dot theory. 
\end{abstract}
{\textit{Keywords}:} quantum mechanics; position-dependent mass; exact solutions; quantum dots; atomic physics.


\section{Introduction}

Models for particles with nontrivial mass distributions have many uses, particularly for explaining electronic and optical properties of semiconductor devices such as nanowires, quantum dots and double heterostructures of diverse geometries \cite{harrison:valavanis:2016,serra:lipparini:1997,elnabulsi:2020,valenciatorres:etal:2020,elnabulsi:2020b}. Quantum dots have properties intermediate between bulk semiconductors and discrete atoms or molecules with similar spectral features. Semiconductor nanostructures, and solid-state quantum systems in general, can be studied by means of effective position-dependent mass (PDM) carriers stuck in order to simulate complex crystalline atomic effects. This theoretical approach has a long history \cite{wannier:1937,slater:1949,luttinger:kohn:1955,bendaniel:duke:1966,gora:williams:1969,zhu:kroemer:1983,bastard:1992}. 

The study of PDM differential equations has been steadily growing along the years, especially in one-dimensional models \cite{cunha:christiansen:2013,christiansen:cunha:2013,christiansen:cunha:2014,lima:christiansen:2022,dacosta:gomez:portesi:2020,ho:roy:2019,schmidt:dejesus:2018}. In more than one-dimension the literature is less abundant, \emph{e.g.} \cite{chang-ying:zhong-zhou:guo-xing:2005,eleuch:jha:rostovtsev:2012,guo-xing:yang:zhong-zhou:2006}.

The PDM problem is basically the fact that the ordinary Schrödinger differential equation becomes one dramatically more complex. Once we consider that the particle mass varies with position \(M\to M(\boldsymbol{r})\) the differential equation suffers three main consequences. The ordinary energy eigenvalue $E$ appears multiplied by a space function, the differential equation acquires new derivative terms and, on top of this, the kinetic energy operator is no longer unique so we may have an ambiguous quantum evolution and spectrum. Since the linear momentum \(\hat{\boldsymbol{p}}\equiv-i\hbar\boldsymbol{\nabla}\) does not commute with the position operator (so neither with the particle mass), several Hamiltonians are possibly defined; this gives rise to the so-called ordering ambiguity {(see, \emph{e.g.}, \cite{vonroos:1983,mustafa:mazharimousavi:2007})}. It is well known that the Schrödinger equation with external potentials has not met many cases with exact analytical solution. Thus, what was already a difficult task becomes much more complex when one considers a variable mass.

In order to handle such Hamiltonians altogether, we adopt a two-parameter hermitian operator \cite{vonroos:1983}. In three dimensions, it reads
\beq
    \hat{H}_{a,b}(\boldsymbol{r})=-\frac{\hbar^2}{4}\left(M^a\hat{\boldsymbol{\nabla}}M^{-1-a-b}\hat{\boldsymbol{\nabla}}M^b
    +M^b\hat{\boldsymbol{\nabla}}M^{-1-a-b}\hat{\boldsymbol{\nabla}}M^a\right)+V(\boldsymbol{r})\;.
\nonumber
\eeq
The parameters \(a,b\in\RR\) are to represent all the possible orderings and \(V(\boldsymbol{r})\) is as usual an external potential. We will treat simultaneously the orderings of BDD (BenDaniel \& Duke) \cite{bendaniel:duke:1966} ($a=b=0$), GW (Gora \& Williams) \cite{gora:williams:1969} ($a=-1$, $b=0$), ZK (Zhu \& Kroemer) \cite{zhu:kroemer:1983} ($a=b=-1/2$), LK (Li \& Kuhn) \cite{li:kuhn:1993} ($a=0$, $b=-1/2$) and MM (Mustafa \& Mazharimousavi) \cite{mustafa:mazharimousavi:2007} ($a=b=-1/4$).


In three dimensions, the kinetic sector of the ordering family of Hamiltonians is
$$\hat{K}_{a,b}(\boldsymbol{r})=\frac{1}{2M}\hat{\boldsymbol{p}}^2-\frac{1}{2M}\left(\frac{1}{M}\hat{\boldsymbol{p}}M\right)\cdot\hat{\boldsymbol{p}}+\frac{1}{2M}\left[\frac{a+b}{2}\hat{\boldsymbol{p}}\left(\frac{1}{M}\hat{\boldsymbol{p}}M\right)-
  \left(ab+\frac{a+b}{2}\right){\left(\frac{1}{M}\hat{\boldsymbol{p}}M\right)}^2\right] ,$$
with \(M\equiv M(\boldsymbol{r})\).
The wave function \(\Psi(\boldsymbol{r})\) thus obeys the following differential equation
\beq
	-{\boldsymbol{\nabla}}^2\Psi(\boldsymbol{r})+\frac{\boldsymbol{\nabla}M}{M}\cdot\boldsymbol{\nabla}\Psi(\boldsymbol{r})+\frac{2M}{\hbar^2}\left(V(\boldsymbol{r})+U_{a,b}(\boldsymbol{r})\right)\Psi(\boldsymbol{r})=\frac{2M}{\hbar^2}E\Psi(\boldsymbol{r}) \;, 
\eeq	
where \(U_{a,b}(\boldsymbol{r})\) contains all the ordering information in terms of  \(a,b\) 
\beq
U_{a,b}(\boldsymbol{r})=-\frac{\hbar^2}{2M}\left[\frac{a+b}{2}\boldsymbol{\nabla}\cdot\left(\frac{\boldsymbol{\nabla}M}{M}\right)-\left(ab+\frac{a+b}{2}\right){\left(\frac{\boldsymbol{\nabla}M}{M}\right)}^2\right].
\eeq
{This potential \(U_{a,b}(\boldsymbol{r})\) emerges independently of the presence of an external potential $V(\boldsymbol{r})$; it is just a direct consequence of considering a position-dependent mass.}

By means of a scale transformation \(\boldsymbol{r}\to\epsilon\boldsymbol{r}\), where \(\epsilon>0\) carries space units which make \(\boldsymbol{r}\) dimensionless, we introduce \(\psi(\boldsymbol{r})\) and \(m(\boldsymbol{r})\equiv m\) such that \(\Psi(\boldsymbol{r})\to\epsilon^{-D/2}\psi(\boldsymbol{r})\) and \(M(\boldsymbol{r})\to m_0m\), where $D$ is the number of spatial dimensions and $m_0>0$ has mass units. {Note the factor \(\epsilon^{-D/2}\) guaranties the normalization condition invariant, \emph{i.e.}, \(1=\int_{\RR^D}{\sqabs{\Psi(\boldsymbol{r})}\text{d}^D\boldsymbol{r}}=\int_{\RR^D}{\sqabs{\psi(\boldsymbol{r})}\text{d}^D\boldsymbol{r}}\).} We also define the dimensionless energy functions 
$\tilde{E}\equiv\frac{2\epsilon^2m_0}{\hbar^2}E,\, \tilde{V}\left(\boldsymbol{r}\right)\equiv\frac{2\epsilon^2m_0}{\hbar^2}V\left(\epsilon\boldsymbol{r}\right)$ and $\tilde{U}_{a,b}\left(\boldsymbol{r}\right)\equiv
	\frac{2\epsilon^2m_0}{\hbar^2}U_{a,b}\left(\epsilon\boldsymbol{r}\right).$
With this, the generalized Schrödinger equation above can be written in a simpler dimensionless form
\beq
 \label{eq:1}
 -\frac{1}{m}{\boldsymbol{\nabla}}^2\psi\left(\boldsymbol{r}\right)+\frac{1}{m}\frac{\boldsymbol{\nabla}m}{m}\cdot\boldsymbol{\nabla}\psi\left(\boldsymbol{r}\right)+\left\{-\frac{1}{m}\left[\frac{a+b}{2}\boldsymbol{\nabla}\cdot\left(\frac{\boldsymbol{\nabla}m}{m}\right)-\left(ab+\frac{a+b}{2}\right){\left(\frac{\boldsymbol{\nabla}m}{m}\right)}^2\right]+\tilde{V}\left(\boldsymbol{r}\right)\right\}\psi\left(\boldsymbol{r}\right)= \tilde{E}\psi\left(\boldsymbol{r}\right).
\eeq
We will solve this equation in the next sections. 


\section{Analytical  solution}

Since we aim to consider a spherical quantum dot, we proceed to separate the differential equation \eqref{eq:1} using spherical coordinates \((r,\theta,\phi)\) such that \(x=r\cos{\phi}\sin{\theta}\), \(y=r\sin{\phi}\sin{\theta}\) and \(z=r\cos{\phi}\), where $r>0$, \(0\leq\phi\leq2\pi\) and \(0\leq\theta\leq\pi\), and factorizing the wave function by \(\psi(\boldsymbol{r})=R(r)\Upsilon(\theta,\phi)\). 
As usual, we obtain an angular differential equation
\beq
    \left[\frac{1}{\sin{\theta}}\pdiff{}{\theta}\left(\sin{\theta}\pdiff{}{\theta}\right)+\frac{1}{\sin^2{\theta}}\pddiff{}{\phi}\right]\Upsilon(\theta,\phi)
    =-\ell(\ell+1)\Upsilon(\theta,\phi) \;,\nonumber
\eeq 
where
\begin{equation}
    \label{eq:3}
    \Upsilon_\ell^\mathscr{m}(\theta,\phi)=\vartheta_\mathscr{m}\sqrt{\frac{2\ell+1}{4\pi}\frac{(\ell-|\mathscr{m}|)!}{(\ell+|\mathscr{m}|)!}}e^{i\mathscr{m}\phi}P_\ell^\mathscr{m}(\cos{\theta}) \quad \text{and} \quad
  \vartheta_\mathscr{m}\equiv
    \begin{cases}
        (-1)^\mathscr{m} &  \mathscr{m}\geq0 \\
        1                &  \mathscr{m}<0
    \end{cases} \;,  
\end{equation}
Here \(\ell\in\ZZ_+\) and \(\mathscr{m}\in\ZZ\) (\(|\mathscr{m}|\leq\ell\)) are respectively the angular and magnetic quantum numbers (note the calligraphic \(\mathscr{m}\) instead of the mathtype \(m\) of the mass), and \(P_\ell^\mathscr{m}(\omega)\) are the Legendre polynomials,
see e.g. \cite[p.~333]{shankar:2013}.

\begin{figure}
	\subfigure[\label{fig:1a}]{\includegraphics[width=.48\linewidth]{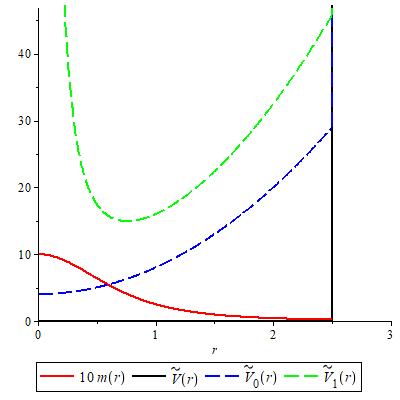}} \ \
	\subfigure[\label{fig:1b}]{\includegraphics[width=.48\linewidth]{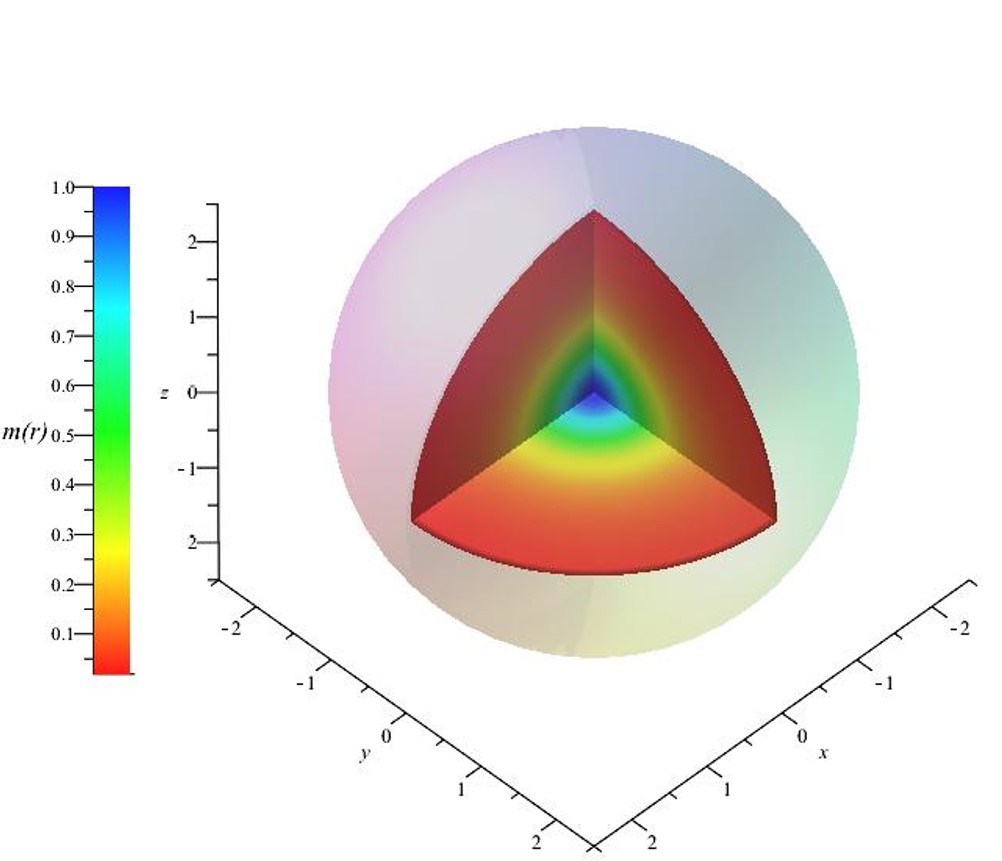}} \ \ 
	\subfigure[\label{fig:1c}]{\includegraphics[width=.48\linewidth]{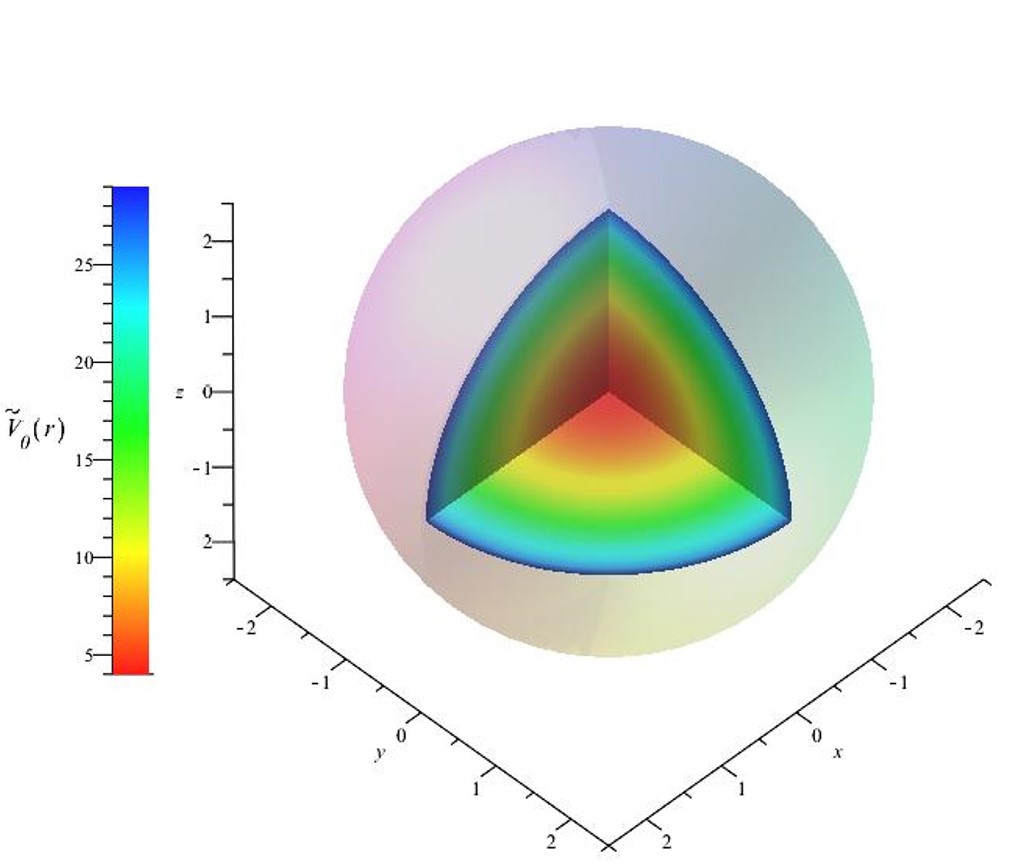}} \ \ 
	\subfigure[\label{fig:1d}]{\includegraphics[width=.48\linewidth]{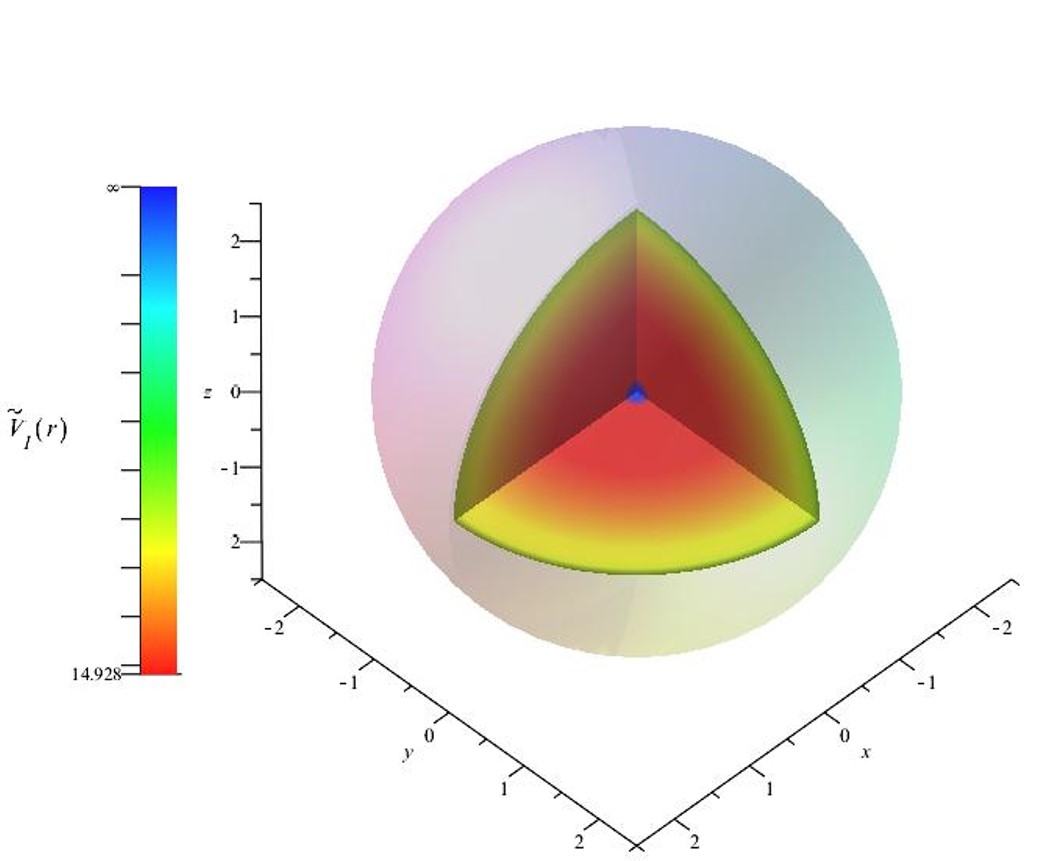}}
	\caption{The position-dependent mass and some radial effective potentials for \(\delta=2.5\) in a spherical quantum dot. Figs.~\subref{fig:1b}, \subref{fig:1c} and \subref{fig:1d} are for the three-dimensional mass (magnified for the sake of resolution), $\tilde{V}_0\left(\boldsymbol{r}\right)$ and $\tilde{V}_1\left(\boldsymbol{r}\right)$ respectively. }
	\label{fig:1}
\end{figure}

The radial differential equation, on the other hand, is a generalized extended case
\begin{equation}
    \label{eq:4}
    -\frac{1}{m}{\left(rR(r)\right)}''+\frac{m'}{m^2}{\left(rR(r)\right)}'+\left(\tilde{U}_{a,b}(r)+\tilde{V}_\ell(r)\right)rR(r)=\tilde{E}\,rR(r)\;,
\end{equation}
which we shall study in what follows. Let us signal important differences with the one-dimensional PDM case, 

\begin{itemize}
    \item[(i)] the kinetic-potential has a new ambiguity-term highly depending on the position
    \beq
        \tilde{U}_{a,b}(r)=-\frac{1}{m(r)}\left[\frac{a+b}{2}{\left(\frac{m'(r)}{m(r)}\right)}'-\left(ab+\frac{a+b}{2}\right){\left(\frac{m'(r)}{m(r)}\right)}^2+\frac{a+b}{r}\left(\frac{m'(r)}{m(r)}\right)\right] \;;
    \eeq
    \item[(ii)] the radial effective potential reads (see Fig.~\ref{fig:1})
    \beq
        \tilde{V}_\ell(r)\equiv\tilde{V}(r)-\frac{1}{m(r)\, r}\frac{m'(r)}{m(r)}+\frac{\ell(\ell+1)}{m(r)\, r^2}\;,
    \eeq
    where  the usual three-dimensional orbital ``\(\ell\)-term'' is now divided by the function $m(r)$ and the ``$m$-term'' is completely new. 
\end{itemize}



In order to solve eq.~\eqref{eq:4} we carry out a point canonical transformation
\begin{subequations}
    \label{eq:5}
    \begin{equation}
        \label{eq:5a}
        \diff{\xi}{r}=m^{1/2} \;,
    \end{equation}
    \begin{equation}
        \label{eq:5b}
        r\,R(r)=m^{1/4}\,\Xi(\xi) \;,
    \end{equation}
\end{subequations}
which leads to an ordinary Schrödinger equation with a highly modified effective potential,
\begin{subequations}
    \label{eq:6}
    \begin{equation}
        \label{eq:6a}
        -\Xi''(\xi)+V_{a,b;\ell}(\xi)\,\Xi(\xi)=\tilde{E}\,\Xi(\xi) \;,
    \end{equation}
The potential term depends on the mass function and its derivatives in a nontrivial way, and on the specific ordering choice of the kinetic Hamiltonian
    \beq
        \label{eq:6b}
        V_{a,b;\ell}(\xi)\equiv\tilde{V}(r(\xi))+\frac{1}{m}\left[\frac{\ell(\ell+1)}{{r(\xi)}^2}
        -\frac{1}{2}\left(a+b+\frac{1}{2}\right){\left(\frac{m'}{m}\right)}'+\left(ab+\frac{a+b}{2}+\frac{3}{16}\right){\left(\frac{m'}{m}\right)}^2-\frac{a+b+1}{r(\xi)}\left(\frac{m'}{m}\right)\right] 
    \eeq
\end{subequations}
where $m=m(r(\xi))$ and the derivative $' = d/dr(\xi)$.

\section{Mass distribution and boundary conditions}  

 It is well known that  quantum dots and few-electron transistors have definite atom-like features such as  
quantized energy spectra \cite{PRL:1988,kastner:1993,ashoori:1996}.  In order to apply our model calculations we can think of an effective carrier confined in a nanometric region. It could be for instance an electron in a semiconductor or a metal nanoparticle surrounded by an insulator or a very strong electric field, e.g. a layer of GaAs surrounded by an insulator such as AlGaAs at low temperatures \cite{scirep:2018}. The case of a perfect insulator can be modeled by a hard shell potential, and the field influence of the crystal structure on the particle by means of a position dependent mass.
To specify the particle spectrum we will consider some continuous mass function confined in a region of size $\epsilon\delta$
\beq
    \label{eq:2}
     m=\frac{1}{(1+r^2)^2} \quad \text{and} \quad
    V(\boldsymbol{r})=V(r)=
    \begin{cases}
        0       &   r<\delta \\
        \infty  &   r\geq\delta
    \end{cases} \;.
\eeq
{Discrete step-wise mass distributions are of course simpler to handle and rougher approaches to the problem. Continuous position-dependent masses are more subtle to represent the many particle structure and have been extensively used in the literature e.g.\cite{cunha:christiansen:2013,christiansen:cunha:2013,christiansen:cunha:2014,lima:christiansen:2022,dacosta:gomez:portesi:2020,ho:roy:2019,schmidt:dejesus:2018, chang-ying:zhong-zhou:guo-xing:2005,eleuch:jha:rostovtsev:2012,guo-xing:yang:zhong-zhou:2006}. The most general treatment of continuous mass profiles, to the best of our knowledge, has been conducted in Ref.
\cite{lima:christiansen:2022}. There, a full comparison among the most used continuous position-dependent masses is presented in a comprehensive calculation in one-dimensional space. In the present effort, we have also tried with a solitonic mass profile following \cite{cunha:christiansen:2013,christiansen:cunha:2013,christiansen:cunha:2014} but we found that the analytical solution to this configuration was not possible. Thus, we decided to build the spherical quantum dot model using a steep decreasing function in order to represent a rapidly varying mass in a tiny region. This mass has another benefit which is that it has not been previously considered in three dimensions so calculations are rather novel.}

The change of variables \eqref{eq:5a} results in
\begin{equation}
    \label{eq:7}
    \xi=\arctan{r}\Leftrightarrow r=\tan{\xi} \;,
\end{equation}
which yields the effective potential in \(\xi\) space
$$V_{a,b;\ell}(\xi)=\ell(\ell+1)\cot^2{\xi}+A_{a,b;\ell}\tan^2{\xi}+B_{a,b;\ell} \;,$$
where \(A_{a,b;\ell}\equiv\ell(\ell+1)+16ab+10(a+b)+6\) and \(B_{a,b;\ell}\equiv2\ell(\ell+1)+6(a+b)+5\). 
Other mass distributions can be equally considered; see e.g. \cite{cunha:christiansen:2013} for a one-dimensional solitonic-like shape. In Table~\ref{tab:1} we can see the value of these coefficients for every ordering. 

\begin{figure}
    \centering
    \includegraphics[width=\textwidth]{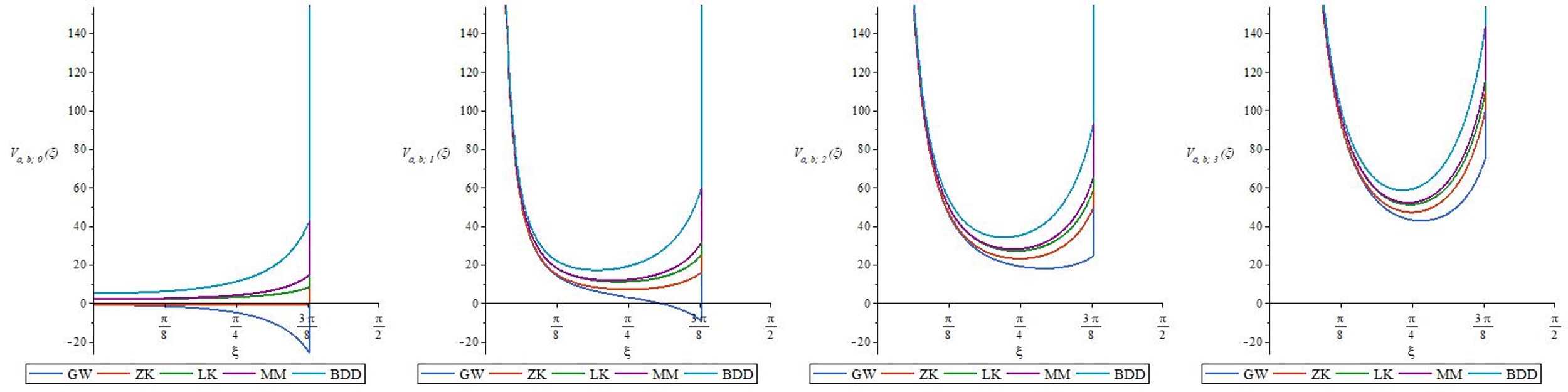}
    \caption{Graphics of \(V_{a,b;\ell}(\xi)\) for \(\delta=2.5\) and some values of \(\ell\).}
    \label{fig:2}
\end{figure}

In Fig.\ref{fig:2} we depict the effective potentials for all the relevant orderings and several orbital numbers.
\begin{longtable}{|c||c|c|c|c|c|}
    \caption{ \(A_{a,b;\ell}\) and \(B_{a,b;\ell}\) for the relevant kinetic orderings.}
    \label{tab:1} \\
    \hline
    \textbf{Ordering}        &   \textbf{GW}         &   \textbf{ZK}         &   \textbf{LK}         &   \textbf{MM}         &   \textbf{BDD} \\ 
    \hline
    \hline
    \boldmath{\(A_{a,b;\ell}\)} &   \(\ell(\ell+1)-4\)  &   \(\ell(\ell+1)\)    &   \(\ell(\ell+1)+1\)  &   \(\ell(\ell+1)+2\)  &   \(\ell(\ell+1)+6\) \\ 
    \hline
    \boldmath{\(B_{a,b;\ell}\)} &   \(2\ell(\ell+1)-1\) &   \(2\ell(\ell+1)-1\) &   \(2\ell(\ell+1)+2\) &   \(2\ell(\ell+1)+2\) &   \(2\ell(\ell+1)+5\) \\ 
    \hline
\end{longtable}

In the limit of a distant spherical shell \(\delta\to\infty\) (in Fig.~\ref{fig:2}, the vertical wall going towards \(\xi=\pi/2\)), the effective potential diverges to \(-\infty\) for any \(A_{a,b;\ell}<0\) (and we must discard the fundamental states \(\ell=0\) and \(\ell=1\) for the GW ordering).



We now proceed by defining \(w=\sin^2{\xi}\), and $W(w)$ such that
$$\Xi(\xi)=w^{\mu_\ell^\pm}{(1-w)}^{\nu_{a,b;\ell}}W(w) \;,$$
where \(\mu_\ell^+=\frac{\ell+1}{2}\), \(\mu_\ell^-=-\frac{\ell}{2}\) and
\(\nu_{a,b;\ell}=\frac{1}{4}\left(1-\sqrt{1+4A_{a,b;\ell}}\right)\).
With this, eq.~\eqref{eq:6a} results in a Gauss hypergeometric equation \cite[Chap.~15]{abramowitz:stegun:1970},
\beq
    \left\{w(1-w)\ddiff{}{w}+\left[\gamma_\ell^\pm-(\alpha_{a,b;\ell}^\pm+\beta_{a,b;\ell}^\pm+1)w\right]\diff{}{w}-\alpha_{a,b;\ell}^\pm\beta_{a,b;\ell}^\pm\right\}W_{a,b;\ell}^\pm(w)=0 \;,
\eeq
where the coefficients are given by
\begin{align*}
    &\alpha_{a,b;\ell}^\pm=\mu_\ell^\pm+\nu_{a,b;\ell}+\frac{1}{2}\sqrt{\Delta_{a,b}} \;, \\
    &\beta_{a,b;\ell}^\pm=\mu_\ell^\pm+\nu_{a,b;\ell}-\frac{1}{2}\sqrt{\Delta_{a,b}} \quad \text{and} \\
    &\gamma_\ell^\pm\equiv2\mu_\ell^\pm+\frac{1}{2}=1\pm\left(\ell+\frac{1}{2}\right)\Leftrightarrow
    \begin{cases}
        \gamma_\ell^+=\frac{3}{2}+\ell \\
        \gamma_\ell^-=\frac{1}{2}-\ell
    \end{cases} \;,
\end{align*}
with \(\Delta_{a,b}\equiv\tilde{E}+16ab+4(a+b)+1\). Since \(\gamma_\ell^\pm\notin\ZZ\), for each \(\mu\) the two linearly independent solutions around $w=0$ are
\begin{align*}
W_{a,b;\ell}^\pm&(w)=\hypergeom{\alpha_{a,b;\ell}^\pm,\beta_{a,b;\ell}^\pm;\gamma_\ell^\pm;w} 
\quad \text{and} \\
    \overline{W}_{a,b;\ell}^\pm(w)&=w^{1-\gamma_\ell^\pm}\, \hypergeom{\alpha_{a,b;\ell}^\pm+1-\gamma_\ell^\pm,\beta_{a,b;\ell}^\pm+1-\gamma_\ell^\pm;2-\gamma_\ell^\pm;w} \;.
\end{align*}
In \(\xi\) coordinates they read
\beq
    \label{eq:8}
    \Xi_{a,b;\ell}^\pm(\xi)={\left(\sin{\xi}\right)}^{2\mu_\ell^\pm}{\left(\cos{\xi}\right)}^{2\nu_{a,b;\ell}}
    \,\hypergeom{\alpha_{a,b;\ell}^\pm,\beta_{a,b;\ell}^\pm;\gamma_\ell^\pm;\sin^2{\xi}} \;.
\eeq
The general solution, \(\Xi_{a,b;\ell}(\xi)\), given by the linear combination of the solutions $\Xi_{a,b;\ell}^\pm(\xi)$ eq.~\eqref{eq:8}, must be consistent with restriction \(\displaystyle \lim_{r\to0^+}{rR(r)}=0\). Considering eqs.~\eqref{eq:2} and \eqref{eq:5} it is equivalent to make \(\Xi_{a,b;\ell}(\xi)\to0\) for \(\xi\to0^+\), which excludes \(\Xi_{a,b;\ell}^-(\xi)\) since it does not converge to zero at the origin.
Therefore, \(\Xi_{a,b;\ell}(\xi)=\mathfrak{C}_{a,b;\ell}\Xi_{a,b;\ell}^+(\xi)\), where \(\mathfrak{C}_{a,b;\ell}\in\CC\) is determined by normalization.

In ordinary radial coordinates, $r$, the radial solutions inside the shell can be obtained by means of eqs.~\eqref{eq:2}, \eqref{eq:5b} and \eqref{eq:7}:
\beq
    \label{eq:9}
    R_{a,b;\ell}(r)=\frac{\mathfrak{C}_{a,b;\ell}r^\ell}{{(1+r^2)}^{\frac{\ell}{2} +\nu_{a,b;\ell}+1}}\;\hypergeom{\alpha_{a,b;\ell}^+,\beta_{a,b;\ell}^+;\frac{3}{2}+\ell;\frac{r^2}{1+r^2}} .
\eeq


\subsection{The energy spectrum}
The energy spectrum \({\left\{\tilde{E}_{a,b;\ell,\mathscr{n}}\right\}}_{\mathscr{n}\in\ZZ_+}\) is obtained by imposing the boundary condition \(R_{a,b;\ell}(\delta)=0\), \(\tilde{E}_{a,b;\ell,\mathscr{n}}\) being the (\(\mathscr{n}+1\))-th real solution of the transcendental equation
\begin{equation}
    \label{eq:10}
    \hypergeom{\alpha_{a,b;\ell}^+,\beta_{a,b;\ell}^+;\frac{3}{2}+\ell;\frac{\delta^2}{1+\delta^2}}=0 \;.
\end{equation}
The eigenstates (orbitals) are denoted by the triple \((\ell,\mathscr{m},\mathscr{n})\); note that eigenstates \((\ell,-\ell,\mathscr{n})\), ..., \((\ell,\ell,\mathscr{n})\) are degenerated in energy.
%
In order to plot our results and to allow numerical comparisons we will adopt \(\delta=2.5\) while leaving free the parameters $m_0$ and \(\epsilon\). 
In Table~\ref{tab:2} we can see the sixteen lowest quantum numbers energy bound states for each ordering.

\begin{figure}
    \centering
    \subfigure[\label{fig:3a}]{\includegraphics[height=.22\textwidth, width=.32\textwidth]{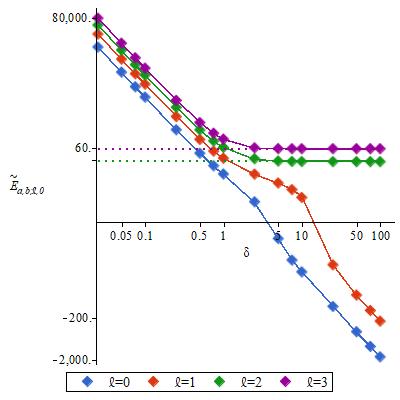}} \ \
    \subfigure[\label{fig:3b}]{\includegraphics[height=.22\textwidth, width=.32\textwidth]{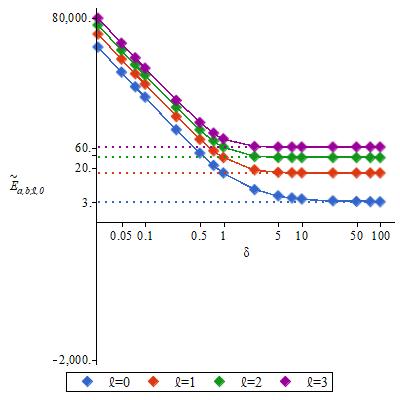}} \ \
    \subfigure[\label{fig:3c}]{\includegraphics[height=.22\textwidth, width=.31\textwidth]{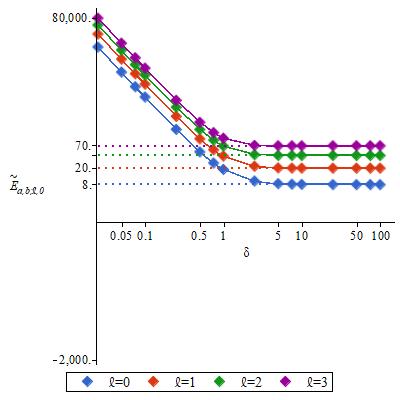}} \ \
    \subfigure[\label{fig:3d}]{\includegraphics[height=.22\textwidth, width=.31\textwidth]{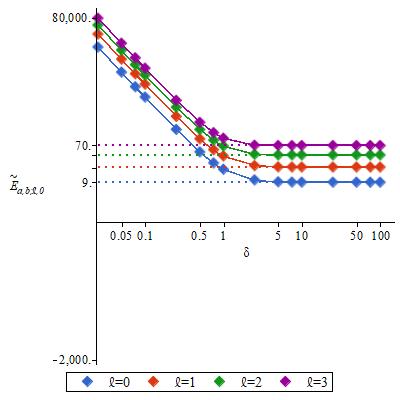}} \ \
    \subfigure[\label{fig:3e}]{\includegraphics[height=.22\textwidth, width=.31\textwidth]{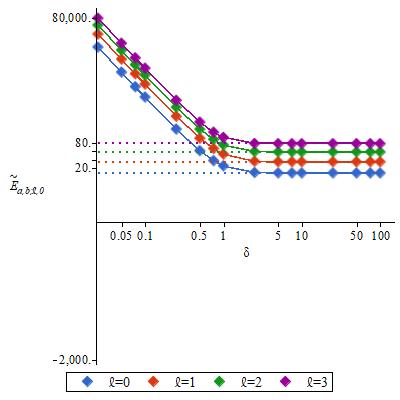}} \ \
    \caption{\(\tilde{E}_{a,b;\ell,0}\) for the orderings \subref{fig:3a} GW, \subref{fig:3b} ZK, \subref{fig:3c} LK, \subref{fig:3d} MM and \subref{fig:3e} BDD, with \(\delta\) varying between $0.025$ and $100$.}
    \label{fig:3}
\end{figure}

\begin{figure}
    \subfigure[\label{fig:4a}]{\includegraphics[height=.22\textwidth,width=.32\linewidth]{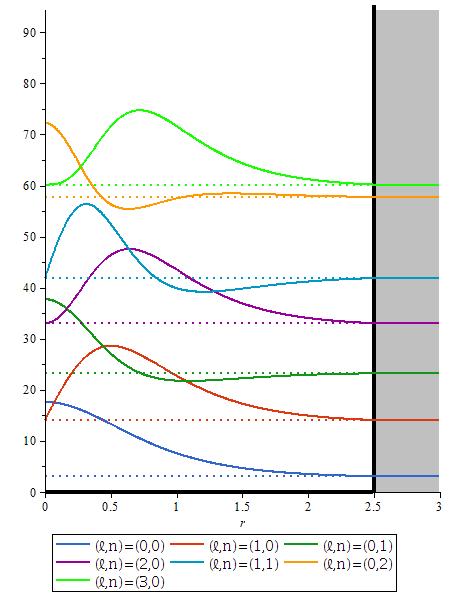}} \ \
    \subfigure[\label{fig:4b}]{\includegraphics[height=.22\textwidth,width=.32\linewidth]{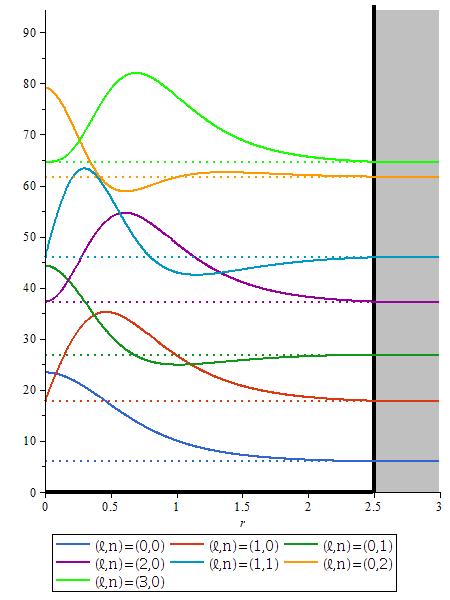}} \ \
    \subfigure[\label{fig:4c}]{\includegraphics[height=.22\textwidth,width=.31\linewidth]{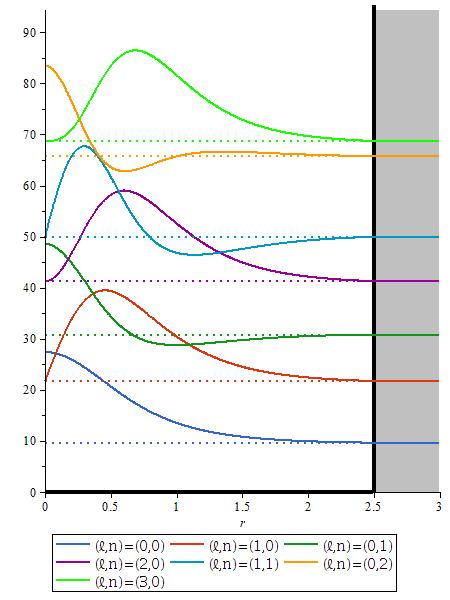}} \ \
    \subfigure[\label{fig:4d}]{\includegraphics[height=.22\textwidth,width=.31\linewidth]{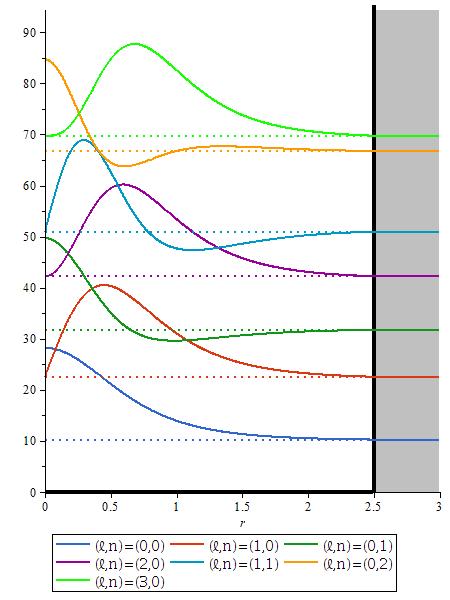}} \ \
    \subfigure[\label{fig:4e}]{\includegraphics[height=.22\textwidth,width=.31\linewidth]{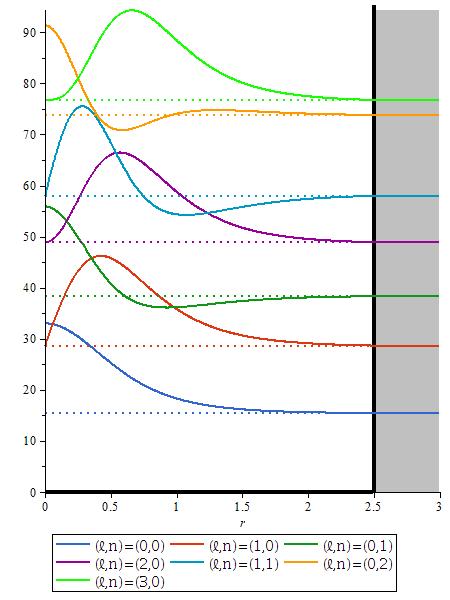}} \ \
    \caption{Radial solutions for the first seven bound-energy levels for the ordering \subref{fig:4a} GW, \subref{fig:4b} ZK, \subref{fig:4c} LK, \subref{fig:4d} MM and \subref{fig:4e} BDD, with \(\delta=2.5\). The solid black line represents the hard wall.}
    \label{fig:4}
\end{figure}

We can obtain more information about these low lying states, for each \(\ell\) and each ordering, by varying \(\delta\). See Fig.~\ref{fig:3}.
In Fig.~\ref{fig:4} we show the graphics of the first radial solutions.

\begin{longtable}{|c|c||c|c|c|c||}
    \caption{Energy levels for the different orderings with \(\ell,\mathscr{n}=0,1,2,3\).}
    \label{tab:2} \\
    \cline{3-6}
        \multicolumn{2}{l}{}                                                        &   \boldmath{\(\ell=0\)}   &   \boldmath{\(\ell=1\)}   &   \boldmath{\(\ell=2\)}   &   \boldmath{\(\ell=3\)}   \\ 
        \hline
        \multirow{4}{*}{\small\textbf{Ordering GW}}   &   \boldmath{\(\mathscr{n}=0\)}    &   $3.0466$                &   $14.049$                &   $33.003$                &   $60.137$                \\  
                                                &   \boldmath{\(\mathscr{n}=1\)}    &   $23.197$                &   $41.829$                &   $67.859$                &   $101.73$                \\  
                                                &   \boldmath{\(\mathscr{n}=2\)}    &   $57.716$                &   $83.566$                &   $116.63$                &   $157.28$                \\  
                                                &   \boldmath{\(\mathscr{n}=3\)}    &   $106.33$                &   $139.26$                &   $179.33$                &   $226.81$                \\  
        \hline
        \multirow{4}{*}{\small\textbf{Ordering ZK}}   &   \boldmath{\(\mathscr{n}=0\)}    &   $5.9662$                &   $17.817$                &   $37.229$                &   $64.597$                \\  
                                                &   \boldmath{\(\mathscr{n}=1\)}    &   $26.865$                &   $45.900$                &   $72.301$                &   $106.48$                \\  
                                                &   \boldmath{\(\mathscr{n}=2\)}    &   $61.695$                &   $87.773$                &   $121.10$                &   $162.00$                \\  
                                                &   \boldmath{\(\mathscr{n}=3\)}    &   $110.46$                &   $143.54$                &   $183.79$                &   $231.47$                \\  
    	\hline
        \multirow{4}{*}{\small\textbf{Ordering LK}}   &   \boldmath{\(\mathscr{n}=0\)}    &   $9.6083$                &   $21.682$                &   $41.222$                &   $68.660$                \\  
                                                &   \boldmath{\(\mathscr{n}=1\)}    &   $30.781$                &   $49.915$                &   $76.404$                &   $110.65$                \\  
                                                &   \boldmath{\(\mathscr{n}=2\)}    &   $65.701$                &   $91.833$                &   $125.22$                &   $166.18$                \\  
                                                &   \boldmath{\(\mathscr{n}=3\)}    &   $114.50$                &   $147.61$                &   $187.91$                &   $235.64$                \\  
    	\hline
        \multirow{4}{*}{\small\textbf{Ordering MM}}   &   \boldmath{\(\mathscr{n}=0\)}    &   $10.220$                &   $22.519$                &   $42.192$                &   $69.704$                \\  
                                                &   \boldmath{\(\mathscr{n}=1\)}    &   $31.695$                &   $50.926$                &   $77.503$                &   $111.82$                \\  
                                                &   \boldmath{\(\mathscr{n}=2\)}    &   $66.710$                &   $92.896$                &   $126.34$                &   $167.37$                \\  
                                                &   \boldmath{\(\mathscr{n}=3\)}    &   $115.55$                &   $148.70$                &   $189.04$                &   $236.81$                \\  
    	\hline
        \multirow{4}{*}{\small\textbf{Ordering BDD}}  &  \boldmath{\(\mathscr{n}=0\)}    &   $15.411$                &   $28.618$                &   $48.849$                &   $76.693$                \\  
                                                &   \boldmath{\(\mathscr{n}=1\)}    &   $38.282$                &   $57.925$                &   $84.848$                &   $119.44$                \\  
                                                &   \boldmath{\(\mathscr{n}=2\)}    &   $73.765$                &   $100.17$                &   $133.85$                &   $175.11$                \\  
                                                &   \boldmath{\(\mathscr{n}=3\)}    &   $122.77$                &   $156.05$                &   $196.56$                &   $244.53$                \\  
    	\hline
\end{longtable}

\subsection{Complete eigenstates}

\begin{figure}
    \subfigure[\label{fig:5a}]{\includegraphics[width=.47\linewidth]{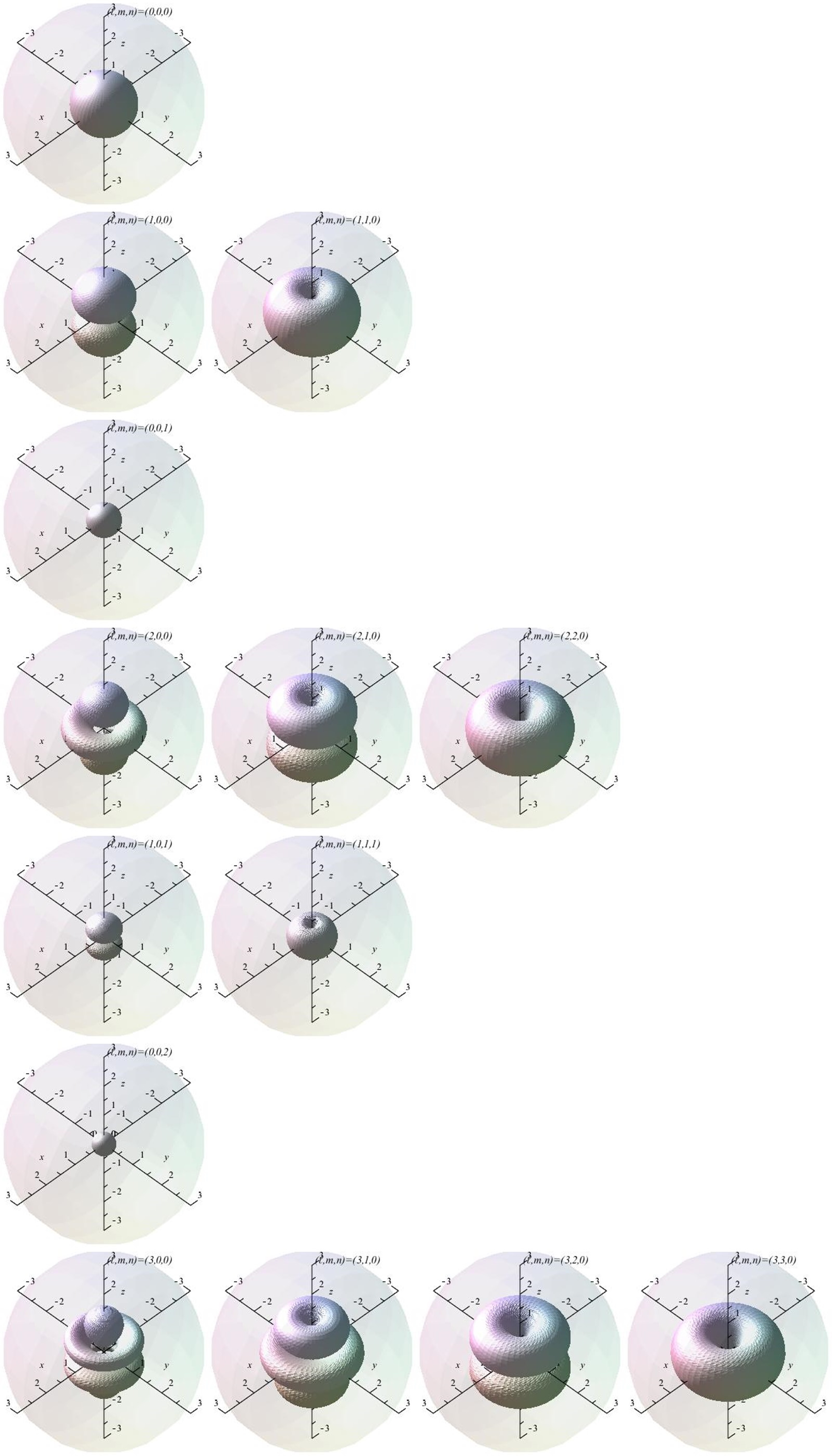}}
    \subfigure[\label{fig:5b}]{\includegraphics[width=.47\linewidth]{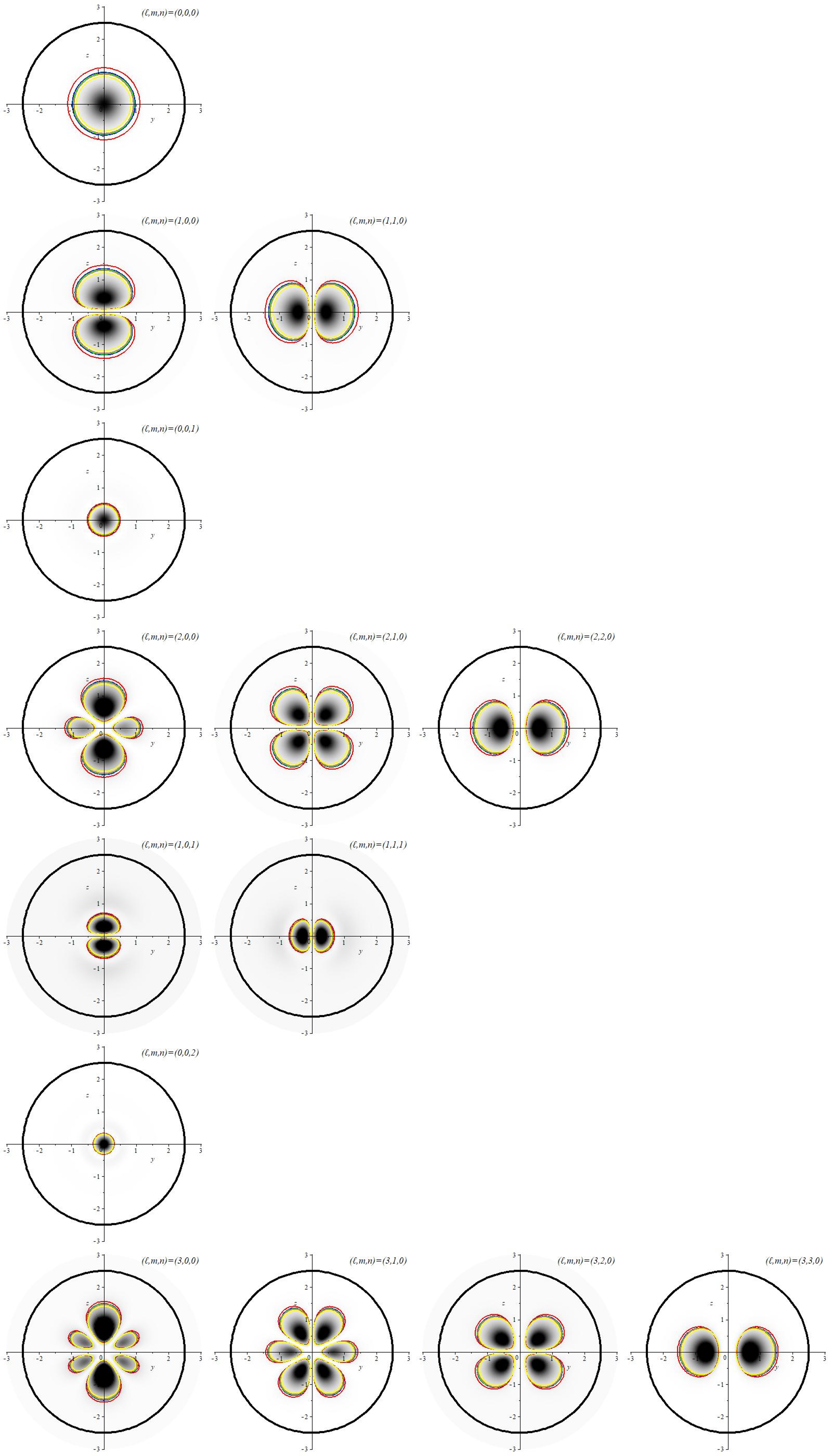}}
    \caption{Three-dimensional particle probability densities \({\rho}_{\ell,\mathscr{m},\mathscr{n}}\) for the first seven lowest energy levels (the darker shade signals the bigger probability density). \subref{fig:5a} {3D confined particle density distributions \({\rho}_{\ell,\mathscr{m},\mathscr{n}}=cons\)}. \subref{fig:5b} {Contour lines for some particular value of \(\rho_{a,b;\ell,\mathscr{m},\mathscr{n}}(\boldsymbol{r})\)  corresponding to the orderings GW (red), ZK (blue), LK (green), MM (golden) and BDD (yellow). The black circumference is for the confining shell.} }
    \label{fig:5}
\end{figure}

The final expressions for the complete eigenstates are obtained by taking together eqs.~\eqref{eq:3}, \eqref{eq:9} and \eqref{eq:10}, resulting, for \(r<\delta\),
\beq
    \label{eq:11}
    \psi_{a,b;\ell,\mathscr{m},\mathscr{n}}(\boldsymbol{r})=\tilde{\mathfrak{C}}_{a,b;\ell,\mathscr{m},\mathscr{n}}e^{i\mathscr{m}\phi}P_\ell^\mathscr{m}(\cos{\theta})\;\dfrac{r^\ell}{{(1+r^2)}^{\frac{\ell}2+\nu_{a,b;\ell}+1}}\hypergeom{\alpha_{a,b;\ell,\mathscr{n}}^+,\beta_{a,b;\ell,\mathscr{n}}^+;\dfrac{3}{2}+\ell;\dfrac{r^2}{1+r^2}} ,
\eeq
and $0\,$ otherwise. Here $\tilde{\mathfrak{C}}_{a,b;\ell,\mathscr{m},\mathscr{n}}\equiv\vartheta_\mathscr{m}\sqrt{\frac{2\ell+1}{4\pi}\frac{\left(\ell-|\mathscr{m}|\right)!}{\left(\ell+|\mathscr{m}|\right)!}}\mathfrak{C}_{a,b;\ell}$, with $\mathfrak{C}_{a,b;\ell}\equiv{\left(\int_0^\delta{\sqabs{R_{a,b;\ell}(r)}r^2\text{d}r}\right)}^{-1/2}$. From eq.~\eqref{eq:11} we compute the probability density of the particle in three-dimensional space
$$\rho_{a,b;\ell,\mathscr{m},\mathscr{n}}(\boldsymbol{r})\equiv\sqabs{\psi_{a,b;\ell,\mathscr{m},\mathscr{n}}(\boldsymbol{r})}=\sqabs{R_{a,b;\ell,\mathscr{n}}(r)}\sqabs{\Upsilon_\ell^\mathscr{m}(\theta,\phi)} \;.$$
 In Fig. \ref{fig:5a} we plot the BDD ordering density \({\rho}_{\ell,\mathscr{m},\mathscr{n}}\equiv\sqabs{{\psi}_{0,0;\ell,\mathscr{m},\mathscr{n}}}\) for it is the most frequently adopted in the PDM literature (the others are similar in shape). In Fig.~\ref{fig:5b} we show all the five orderings contour-lines for constant probability density to let clear that they are all distinct. The individual orbital graphics are organized in a growing order of energy values (horizontally, we display all the degenerated orbitals).



\section{Final comments}

The energy levels of a particle in a quantum dot can be computed using a three-dimensional shell model.
The quantum dot absorption and emission frequencies correspond to transitions between discrete energy levels in this cell and are analogous to atomic spectra. Quantum dots are referred to as artificial atoms whereas electronic wave functions  resemble the ones in real atoms \cite{nature:1999}.
In this paper we have obtained the  analytical solutions \textcolor{blue}{for} a three-dimensional position-dependent mass generalized Schrödinger equation with spherical symmetry confined in three dimensional space for five different kinetic orderings. Once boundary conditions are fixed, our exact analytical expressions depend on three quantum numbers (\(\ell\), \(\mathscr{m}\), \(\mathscr{n}\)) and two ordering parameters $a$ and $b$. We selected a smooth continuous mass profile of inverse biquadratic shape and hard shell boundary conditions which could represent effective electron carriers in a semiconductor surrounded by high vacuum or a good dielectric medium. 
 We computed the eigenfunctions and probability densities of the first sixteen eigenstates (namely \(\ell=0,1,2,3\) and \(\mathscr{n}=0,1,2,3\)), for each of the five ordering levels, BDD, GW, ZK, LK and MM.

We have graphically shown the distinction among all the ordering wavefunctions together with the analytical expressions to reveal their exact differences, see eq.~\eqref{eq:10} (\emph{cf.} Fig.~\ref{fig:4}). We found that the energy levels grow according to \(GW \to ZK \to LK \to MM \to BDD\) (\emph{cf.} Table~\ref{tab:2}). These features will eventually determine the best spectral fit to a given quantum dot experimental outcome and then the corresponding ordering choice to modeling such material.
In order to further sharpen differences among orderings, we also examined the spectral behavior upon variation of \(\delta\) (\emph{cf.} Fig.~\ref{fig:3}). We noted
\begin{itemize}
    \item[(i)] for decreasing \(\delta\) the energy levels grow in an approximately exponential way;
    \item[(ii)] for growing \(\delta\) the energy levels show two possible behaviors: diverge to \(-\infty\) (when \(A_{a,b;\ell}<0\)), 
    or tend to a finite limit which depends on the value of the ordering parameters $a$, $b$ and on \(\ell\) (when \(A_{a,b;\ell}\geq0\)); stability starts to be attained after \(\delta \approx 2.5\).
\end{itemize}
Regarding the last item, when \(\delta\to\infty\) the boundary condition on the radial solution \(R(\delta)=0\) can be replaced by \(\Xi(\pi/2)=0\). For \(A_{a,b;\ell}<0\) it implies a continuum spectrum with no minimal value whereas for \(A_{a,b;\ell}\geq0\) we have \(\hypergeom{\alpha_{a,b;\ell}^+,\beta_{a,b;\ell}^+;\gamma_\ell^+;1}=0\) which yields an analytical expression for the energy levels \cite[eq.~15.1.20]{abramowitz:stegun:1970},
\begin{equation}
    \label{eq:12}
    {\left.\tilde{E}_{a,b;\ell,\mathscr{n}}\right|}_{\delta\to\infty}={\left(2\mathscr{n}+2+\ell-2\nu_{a,b;\ell}\right)}^2-[16ab+4(a+b)+1] \;.
\end{equation}
In Table~\ref{tab:3} one can verify that \(\delta=100\) serves as a good approximation to this asymptotic situation; this is tantamount to a free PDM particle in three dimensions.
\begin{longtable}{|c||c|c|c|c||c|c|c|c|}
    \caption{Energy levels of the lowest  \(\mathscr{n}=0\) states for large \(\delta\) values from eq.~\eqref{eq:10}  and eq.~\eqref{eq:12}.}
    \label{tab:3} \\
    \hline
        \multirow{2}{*}{\textbf{Ordering}}  &   \multicolumn{4}{l|}{\boldmath{\(\delta=100\)}}                                                              & \multicolumn{4}{l|}{\boldmath{\(\delta\to\infty\)}}                                                          \\ 
        \cline{2-9}
                                            &   \boldmath{\(\ell=0\)}   &   \boldmath{\(\ell=1\)}   &   \boldmath{\(\ell=2\)}   &   \boldmath{\(\ell=3\)}   &   \boldmath{\(\ell=0\)}   &   \boldmath{\(\ell=1\)}   &   \boldmath{\(\ell=2\)}   &   \boldmath{\(\ell=3\)}   \\ 
        \hline
        \small\textbf{GW}                         &   -$1732.6$               &   -$240.60$               &   $28.000$                &   $57.351$                &   \(-\infty\)             &   \(-\infty\)	            &   $28.$                   &   $57.351$                \\ 
        \small\textbf{ZK}                         &   $3.0514$                &   $15.000$                &   $35.000$                &   $63.000$                &   $3.$                    &   $15.$                   &   $35.$                   &   $63.$                   \\ 
        \small\textbf{LK}                         &   $7.8547$                &   $19.514$                &   $39.348$                &   $67.260$                &   $7.8539$                &   $19.514$                &   $39.348$                &   $67.260$                \\ 
        \small\textbf{MM}                         &   $9.0000$                &   $20.808$                &   $40.606$                &   $68.474$                &   $9.$                    &   $20.808$                &   $40.606$                &   $68.474$                \\ 
        \small\textbf{BDD}                        &   $15.000$                &   $27.861$                &   $48.000$                &   $75.948$                &   $15.$                   &   $27.862$                &   $48.$                   &   $75.948$                \\ 
        \hline
\end{longtable}
The results presented in this paper are of special interest in the study of spherical quantum dots \cite[p.~117]{agrawal:2013}. 
For a typical nanostructure we can set \(\epsilon\sim10^{-9}\)m in the equations above which makes the conversion factor between \(\tilde{E}\) and $E$ of about $\frac{0.038101}{m_0/m_e}${eV};
$m_e$ is the free electron mass. Depending on the composition of the quantum dot one can estimate $m_0/m_e$ from $0.01$ up to order 100 or even more \cite{dang:neu:romestain:1982,fanciulli:lei:moustakas:1993,miwa:fukumoto:1993,fan:etal:1996,li:etal:1996,long:harrison:hagston:1996,paiva:etal:2002,margaritondo:2005,wasserman:2005}.


\section{Acknowledgement}

The authors would like to thank Conselho Nacional de Desenvolvimento Científico e Tecnológico (CNPq) for partial support.
\\

Data Availability Statement: The authors declare that the data supporting the findings of this study are available within the paper.
\\

Author Contribution Statement: The authors declare that they contributed equally to the research.


\bibliographystyle{ieeetr}
\bibliography{refpdm}

\end{document}